\documentclass[twocolumn,prd,superscriptaddress,altaffilletter,nofootinbib]{revtex4-1}
\usepackage{amsmath}
\usepackage{amssymb}
\usepackage{amsfonts}
\usepackage{comment}
\usepackage{graphicx}
\usepackage{hyperref}
\usepackage{enumerate}
\usepackage[nolist]{acronym}
\usepackage[usenames,dvipsnames]{xcolor}
\usepackage{xspace}
\usepackage{tikz}
\usepackage{ulem}
\usepackage{siunitx}
\usepackage{multirow}
\usepackage{bm}
\usepackage{romannum}
\usepackage{mathtools}

\usetikzlibrary{arrows,shapes,trees,decorations.pathreplacing}
\DeclareSIUnit \parsec {pc}

\allowdisplaybreaks

\newcommand{\fhalf}{f_{1/2}}

\newcommand{\Msun}{M_{\odot}}
\newcommand{\e}{\mathrm{e}}
\newcommand{\iu}{\mathrm{i}}
\newcommand{\Mc}{\mathcal{M}}
\newcommand{\DL}{D_{\mathrm{L}}}

\newcommand{\Korig}{K_{\mathrm{orig}}}
\newcommand{\KMB}{K_{\mathrm{MB}}}
\newcommand{\NL}{N_L}
\newcommand{\df}{\Delta f}
\newcommand{\flow}{f_{\mathrm{low}}}
\newcommand{\fhigh}{f_{\mathrm{high}}}
\newcommand{\dt}{\Delta t}

\newcommand{\tc}{t_{\mathrm{c}}}
\newcommand{\tcmin}{t_{\mathrm{c}, \mathrm{min}}}
\newcommand{\tcmax}{t_{\mathrm{c}, \mathrm{max}}}
\newcommand{\wb}{w^{(b)}}
\newcommand{\fb}{f^{(b)}}
\newcommand{\fbp}{f^{(b+1)}}
\newcommand{\dfb}{\Delta f^{(b)}}
\newcommand{\dfbp}{\Delta f^{(b+1)}}
\newcommand{\dtb}{\Delta t^{(b)}}
\newcommand{\Tb}{T^{(b)}}
\newcommand{\Nb}{N^{(b)}}
\newcommand{\Mb}{M^{(b)}}
\newcommand{\Ks}{K_{\mathrm{s}}^{(b)}}
\newcommand{\Ke}{K_{\mathrm{e}}^{(b)}}
\newcommand{\mmax}{m_{\mathrm{max}}}

\renewcommand{\mod}{\mathrm{mod}}
\newcommand{\ceil}[1]{\left\lceil #1 \right\rceil}
\newcommand{\floor}[1]{\left\lfloor #1 \right\rfloor}

\newcommand{\figref}[1]{Fig.~\ref{#1}}
\newcommand{\secref}[1]{Sec.~\ref{#1}}
\newcommand{\appref}[1]{Appendix \ref{#1}}
\newcommand{\RNum}[1]{\uppercase\expandafter{\romannumeral #1\relax}}

\begin{document}

\title{Accelerating parameter estimation of gravitational waves from compact binary coalescence using adaptive frequency resolutions} 

\author{Soichiro Morisaki}
  \affiliation{Department of Physics, University of Wisconsin-Milwaukee, Milwaukee, WI 53201, USA}
\date{\today}

\begin{abstract}

Bayesian parameter estimation of gravitational waves from compact binary coalescence (CBC) typically requires the generation of millions of computationally expensive template waveforms.
We propose a technique to reduce the cost of waveform generation by exploiting the chirping behavior of CBC signal.
Our technique does not require waveforms at all frequencies in the frequency range used in the analysis, and does not suffer from the fixed cost due to the upsampling of waveforms.
Our technique speeds up the parameter estimation of typical binary neutron star signal by a factor of $\mathcal{O}(10)$ for the low-frequency cutoff of $20\,\si{\hertz}$, and $\mathcal{O}(10^2)$ for $5\,\si{\hertz}$.
It does not require any offline preparations or accurate estimates of source parameters provided by detection pipelines.

\end{abstract}

\maketitle

\section{Introduction} \label{sec:introduction}

The discovery of gravitational waves from binary black hole coalescence opened a new window to the Universe \cite{Abbott:2016blz}.
In 2017, gravitational waves from binary neutron star coalescence were also detected \cite{TheLIGOScientific:2017qsa}.
The multi-messenger observations of this event enabled us to measure the Hubble constant in a way independent from the cosmic ladder \cite{Abbott:2017xzu}, learn the origin of heavy elements \cite{Drout:2017ijr, Kasliwal:2017ngb, Cowperthwaite:2017dyu, Tanvir:2017pws, Utsumi:2017cti, Tanaka:2017qxj} and the structure of ultra-relativistic jet from the merger \cite{Troja:2017nqp, Hallinan:2017woc, Alexander:2017aly, Margutti:2017cjl, Mooley:2017enz}.
So far, several tens of compact binary coalescence (CBC) events have been detected \cite{LIGOScientific:2018mvr, Abbott:2020niy} by the LIGO-Virgo collaboration \cite{Harry:2010zz, TheVirgo:2014hva}, enabling us to learn the population properties of binary black holes \cite{LIGOScientific:2018jsj, Abbott:2020gyp}.

In this era of gravitational-wave astronomy, an accurate inference of the source properties from gravitational-wave data is important.
The LIGO-Virgo collaboration employs Bayesian inference with stochastic sampling \cite{Veitch:2014wba, Ashton:2018jfp}, which generates thousands of random samples following the probability distribution of the source parameters conditioned on data.
It typically requires generation of millions of computationally expensive template waveforms.
The waveform generation becomes very costly for light binaries, whose signals have long durations and extend to high frequencies.
For a typical binary neutron star signal, the parameter estimation can take a few weeks, or even years, without any approximate methods.
To solve this issue, various techniques to speed up the inference have been proposed \cite{Canizares:2014fya, Smith:2016qas, Morisaki:2020oqk, Pankow:2015cra, Lange:2018pyp, Wysocki:2019grj, Vinciguerra:2017ngf, Zackay:2018qdy, Talbot:2019okv, Cornish:2010kf, Cornish:2021wxy, Smith:2019ucc, Gabbard:2019rde, Green:2020hst}.

One of the techniques widely used in the detection and parameter estimation of CBC signal is multi-banding \cite{marion:in2p3-00014163, Buskulic:2010zz, Cannon:2011vi, Vinciguerra:2017ngf}.
It exploits the chirping behavior of CBC signal, whose frequency simply increases with time.
In the time domain, it means the sampling frequency can be lowered at the early stage of inspiral, which significantly reduces the number of time samples at which waveforms are evaluated.
This idea has been utilized to speed up matched filtering of data in the detection of CBC signals \cite{Cannon:2011vi}.

On the other hand, the standard parameter estimation is performed in the frequency domain.
The Fourier transform of a CBC waveform is an oscillating function of frequency, and the frequency scale of the oscillations is the inverse of time-to-merger.
Since the time-to-merger decreases as the frequency increases, we can use coarser frequency resolutions at high frequencies for resolving that oscillatory behavior.
The previous study \cite{Vinciguerra:2017ngf} has proposed an efficient technique, which computes waveforms with coarser frequency resolutions at high frequencies and upsamples them to the original frequency resolution.
This technique was named MB-Interpolation, and it significantly reduces the number of waveform evaluations at high frequencies.

However, the overall speed-up gain of MB-Interpolation is more modest than the reduction of the number of waveform evaluations.
For example, it reduces the number of waveform evaluations by a factor of $\sim 60$ for typical BNS signal and the low-frequency cutoff of $20\,\si{\hertz}$, but the speed-up gain of parameter estimation with the TaylorF2 \cite{Buonanno:2009zt} waveform model is $\sim 3$ (See Table 1 of \cite{Vinciguerra:2017ngf}).
It arises because MB-Interpolation requires the upsampling of waveforms and the computation of inner products of upsampled waveforms and data.
Their costs are proportional to the original frequency samples, and are not reduced by multi-banding.

In this paper, we propose another technique, which exploits the chirping behavior of CBC signal but does not require the upsampling. 
If the length of data is $T$, the inner products of waveforms and data require waveforms at frequency points with the frequency interval of $1/T$.
On the other hand, if $\fhalf$ is the frequency from which the time-to-merger is $T/2$, the inner products in $f>\fhalf$ can be approximately computed with the latter half of data and waveforms at frequency points with the frequency interval of $2/T$.
Generalizing this idea, we divide the total frequency range into multiple bands in a way that the time-to-merger of each band is much smaller than that of the previous band, and use coarser frequency resolutions in high-frequency bands.
It significantly reduces the number of waveform evaluations at high frequencies, and does not require upsampled waveforms.

This paper is organized as follows.
First, we formulate our technique and evaluate its speed-up gains in \secref{sec:formulation}.
Next, we investigate the accuracy of our technique in \secref{sec:validation}.
Finally, we summarize the results and conclude this paper in \secref{sec:conclusion}.

\section{Parameter estimation with adaptive frequency resolutions} \label{sec:formulation}

In this section, we formulate our technique and evaluate its speed-up gains.
First, we review the basics of Bayesian parameter estimation of CBC signal in \secref{sec:basics}.
Next, we formulate our technique in Sec. \ref{sec:introduce_window} -- \ref{sec:dfB}.
Finally, we evaluate the speed-up gains in \secref{sec:speedup}.

\subsection{Bayesian inference} \label{sec:basics}

In the Bayesian inference, the probability distribution of model parameters conditioned on data is calculated via the Bayes' theorem,
\begin{equation}
p(\bm{\theta}|\bm{d}) \propto \pi(\bm{\theta}) \mathcal{L}(\bm{d}|\bm{\theta}),
\end{equation}
where $\bm{d}=\left(d_0, d_1, \dots, d_{N-1}\right)^T$ represents time-domain data at $N$ time samples, whose sampling rate is $1/\dt$, and $\bm{\theta}$ represents the model parameters.
$p(\bm{\theta}|\bm{d})$, $\pi(\bm{\theta})$ and $\mathcal{L}(\bm{d}|\bm{\theta})$ are referred to as posterior, prior and likelihood respectively.
The prior is determined based on our prior knowledge or belief on $\bm{\theta}$.

In the standard parameter estimation of CBC signal, the noise is modeled as stationary Gaussian random process.
In this model, the logarithm of likelihood for a single detector is given by \cite{Veitch:2014wba}
\begin{equation}
\ln \mathcal{L}(\bm{d}|\bm{\theta}) = \left(\bm{d}, \bm{h}(\bm{\theta})\right) - \frac{1}{2} \left(\bm{h}(\bm{\theta}), \bm{h}(\bm{\theta})\right) + \mathrm{const.},
\end{equation}
where $\bm{h}(\bm{\theta})$ is the CBC waveform for $\bm{\theta}$.
The inner products are given by
\begin{align}
&\left(\bm{d}, \bm{h}\right) \equiv \frac{4}{T} \Re\left[\sum_{k=1}^{\floor{(N-1)/2}} \frac{\tilde{d}^\ast_k \tilde{h}(f_k)}{S_k} \right], \label{eq:original_d_h} \\
&\left(\bm{h}, \bm{h}\right) \equiv \frac{4}{T} \sum_{k=1}^{\floor{(N-1)/2}} \frac{\left|\tilde{h}(f_k)\right|^2}{S_k} ,
\end{align}
where $T$ is the duration of data $T \equiv N \dt$, $f_k$ is the frequency of the $k$-th bin $f_k \equiv k / T$, $\floor{x}$ is the greatest integer less than or equal to $x$, $\tilde{d}_k$ is the Fourier component of data defined by
\begin{equation}
\tilde{d}_k \equiv \dt \sum_{m=0}^{N-1} d_m \e^{-2 \pi \iu k m / N},
\end{equation}
$S_k$ is the one-sided power spectral density (PSD) of the detector's noise, and $\tilde{h}(f)$ is the template waveform in the frequency domain. 
The DC and Nyquist frequencies have been excluded, and $\bm{\theta}$ has been omitted for ease of notation.
Typically, the low and high frequency cutoffs, $\flow$ and $\fhigh$, have been determined before the analysis.
Here we assume 
\begin{equation}
S_k=+\infty,~~~(f_k<\flow~~\text{or}~~f_k>\fhigh)
\end{equation}
so that the components outside the frequency range are automatically vanishing.
The likelihood for multiple detectors is the product of likelihood for each detector.

In the inference with stochastic sampling, the non-constant part of log-likelihood, which is often referred to as log-likelihood-ratio,
\begin{equation}
\ln \Lambda (\bm{d}|\bm{\theta}) \equiv \left(\bm{d}, \bm{h}(\bm{\theta})\right) - \frac{1}{2} \left(\bm{h}(\bm{\theta}), \bm{h}(\bm{\theta})\right),
\end{equation}
is computed tens to hundreds of millions of times \cite{Veitch:2014wba, Smith:2019ucc}.
It requires the evaluations of $\tilde{h}(f)$ at $\Korig$ frequency points, where
\begin{equation}
\Korig \equiv \floor{\fhigh T} - \ceil{\flow T} + 1 \sim (\fhigh - \flow) T. \label{eq:Korig}
\end{equation}
Our technique is an approximate method to compute log-likelihood-ratio with fewer waveform evaluations.

\subsection{Window functions} \label{sec:introduce_window}

For dividing the total frequency range into multiple frequency bands, we introduce the following overlapping window functions,
\begin{equation}
\wb (f) = \begin{dcases}
\displaystyle \frac{1}{2} \left( 1 + \cos\left( \pi \frac{f - f^{(b)}}{\df^{(b)}} \right) \right), \\
~~~~~~~~(\fb - \dfb < f < \fb)  \\ 
\displaystyle 1,~~~~~(\fb \leq f \leq \fbp - \dfbp) \\
\displaystyle \frac{1}{2} \left( 1 -  \cos\left( \pi \frac{f - \fbp}{\dfbp} \right)  \right), \\
~~~~~~~~(\fbp - \dfbp < f < \fbp) \\
\displaystyle 0,~~~~~(\text{otherwise})
\end{dcases} \label{eq:window_def}
\end{equation}
where
\begin{equation}
\begin{aligned}
&\flow = f^{(0)} < f^{(1)} < \dots < f^{(B)} = \fhigh + \df^{(B)}, \\
&\fb< \fbp - \dfbp,~~~\df^{(0)}=0,
\end{aligned}
\end{equation}
and $B$ represents the number of frequency bands.
$\df^{(B)}$ needs to be positive to smooth the high-frequency end of waveform.
We explain its necessity and our choice of $\df^{(B)}$ in \secref{sec:dfB}.
The window functions are constructed so that their sum becomes unity,
\begin{equation}
\sum_{b=0}^{B-1} w^{(b)}(f) = 1.~~~(\flow \leq f \leq \fhigh) \label{eq:unity}
\end{equation}
The reason for using smooth window functions rather than rectangular window functions is explained in \appref{sec:choice_of_window}.

\subsection{How to compute $(d, h)$} \label{sec:d_h}

First, we introduce an approximate method to compute $(\bm{d}, \bm{h})$.
It can be rewritten as follows thanks to \eqref{eq:unity},
\begin{equation}
(\bm{d}, \bm{h}) = \sum^{B-1}_{b=0} \frac{4}{T} \Re\left[\sum_{k=1}^{\floor{(\Nb-1)/2}} \wb(f_k) \frac{\tilde{d}^\ast_k \tilde{h}(f_k)}{S_k}\right], \label{eq:divide_df}
\end{equation}
where $\Nb$ is an integer satisfying
\begin{equation}
\floor{\frac{\Nb - 1}{2}} \geq \fbp T.
\end{equation}
We use the minimum power of $2$ satisfying this condition as $\Nb$ for efficient Fourier transforms, which specifically speeds up the IFFT-FFT computations for $(\bm{h}, \bm{h})$ introduced in \secref{sec:h_h}.
If $\Nb>N$, we pad zeros to data so that \eqref{eq:divide_df} is satisfied,
\begin{equation}
\frac{\tilde{d}_k}{S_k}=0.~~~\left( k > \floor{\frac{N-1}{2}} \right).
\end{equation}
The inner product can be transformed into a sum over times as follows,
\begin{equation}
\begin{aligned}
&\frac{4}{T} \Re\left[\sum_{k=1}^{\floor{(\Nb-1)/2}}  \wb(f_k)\frac{ \tilde{d}^\ast_k \tilde{h}(f_k)}{S_k}\right] \\
&=2 \dtb \sum^{\Nb-1}_{m=0} D^{(b)}_m h^{(b)}_m,
\end{aligned}
\end{equation}
where $\dtb \equiv N \dt / \Nb$, and
\begin{align}
&D^{(b)}_m \equiv \frac{2}{T} \Re \left[\sum_{k=1}^{\floor{(\Nb - 1)/2}} \frac{\tilde{d}_k}{S_k} \e^{2 \pi \iu k m / \Nb}\right], \label{eq:Dbm} \\
&h^{(b)}_m \equiv \nonumber \\
&~~\frac{2}{T} \Re \left[\sum_{k=1}^{\floor{(\Nb - 1)/2}}  \wb(f_k) \tilde{h}(f_k) \e^{2 \pi \iu k m / \Nb} \right].
\end{align}

\begin{figure}[t]
        \begin{center}
                \includegraphics[width = \columnwidth]{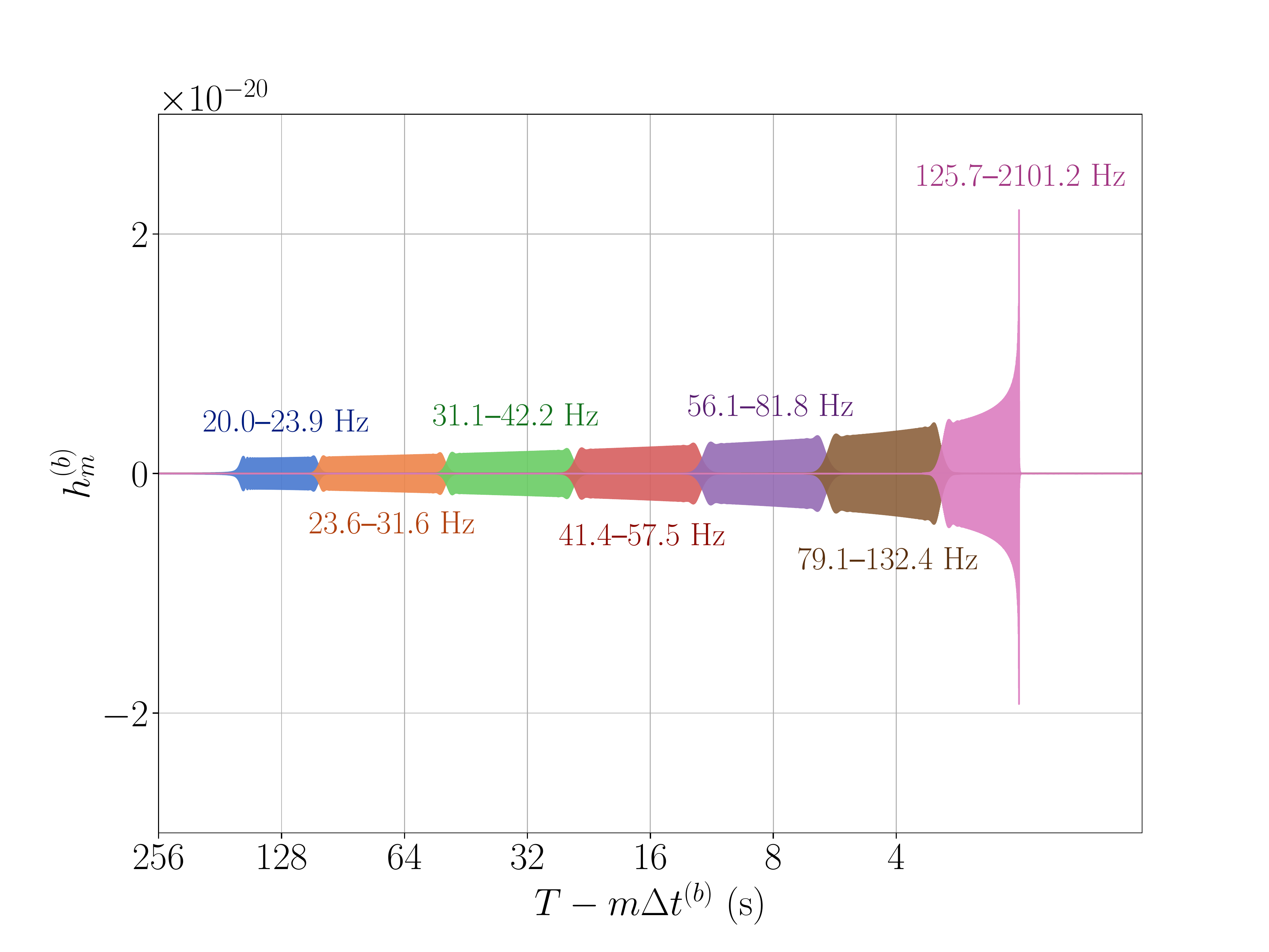}
                \caption{The inverse Fourier transform of the waveform in each frequency band.
                The waveforms are for non-spinning $1.4\Msun\text{--}1.4\Msun$ BNS signal, whose coalescence time is at $T-2\,\si{\second}$.
                The total frequency range starting from $20\,\si{\hertz}$ is divided into $7$ frequency bands, and each band is constructed so that the waveform in the $b$-th band is vanishing except for in the last $2^{8-b}\,\si{\second}$.
                Each label presents the start and end frequencies of the band, $\fb - \dfb$ and $\fbp$. }
                \label{fig:multiband}
        \end{center}
\end{figure}

Since the frequency of the CBC waveform simply increases with time, $h^{(b)}_m$ is almost vanishing for $m \dtb \lesssim T - \tau(\fb - \dfb)$, where $\tau(f)$ is the time to merger from a frequency $f$.
If the waveform contains multiple gravitational-wave moments, $\tau(f)$ is defined on the moment with the maximum magnetic number, whose time to merger is the longest.
Figure \ref{fig:multiband} shows $h^{(b)}_m$ for $1.4\Msun\text{--}1.4\Msun$ BNS with zero spins, which ends at $T-2\,\si{\second}$. 
Each frequency band is constructed so that $\tau(\fb - \dfb) < (2^{8-b}-2)\,\si{\second}$.
The figure shows that $h^{(b)}_m$ is almost vanishing at $m \dt^{(b)} <T - 2^{8-b}\,\si{\second}$, which validates ignoring $h^{(b)}_m$ there.
It implies we can make the approximation that
\begin{equation}
h^{(b)}_m\simeq0,~~~(m=0, 1, \dots, N^{(b)}-M^{(b)}-1) \label{eq:approx1}
\end{equation}
where $\Tb \equiv M^{(b)} \dtb$ is long enough compared to $\tau(\fb - \dfb)$, and
\begin{equation}
T^{(B-1)} < T^{(B-2)} < \dots < T^{(0)} \leq T. \label{eq:Tbs}
\end{equation}
This approximation leads to
\begin{equation}
\sum^{\Nb-1}_{m=0} D^{(b)}_m h^{(b)}_m \simeq \sum^{\Nb-1}_{m=\Nb - \Mb} D^{(b)}_m h^{(b)}_m.
\end{equation}

Finally, the inner product can be transformed into a sum over frequencies as follows,
\begin{align}
&2 \dtb \sum^{\Nb-1}_{m=\Nb - \Mb} D^{(b)}_m h^{(b)}_m \nonumber \\
&\simeq \frac{4}{\Tb} \Re \left[ \sum_{k=\Ks}^{\Ke} \wb(f^{(b)}_k) \tilde{D}^{(b)\ast}_k \tilde{h}(f^{(b)}_k) \right], 
\end{align}
where $\Ks=\ceil{(f^{(b)} - \df^{(b)}) T^{(b)}}$, $\Ke=\floor{f^{(b+1)} T^{(b)}}$, $f^{(b)}_k = k / T^{(b)}$ and
\begin{equation}
\tilde{D}^{(b)}_k = \dtb \sum^{\Nb-1}_{m=\Nb-\Mb} D_m \e^{- 2 \pi \iu k m / \Mb}. \label{eq:tDbm} 
\end{equation}
Here, we have made the following approximation,
\begin{equation}
\dtb \sum^{\Nb-1}_{m=\Nb-\Mb} h^{(b)}_m \e^{- 2 \pi \iu k m / \Mb} \simeq \wb(f^{(b)}_k) \tilde{h}(f^{(b)}_k).
\end{equation}

Finally, the inner product is reduced to
\begin{equation}
\begin{aligned}
&(\bm{d}, \bm{h}) \simeq  \\
&~~~\sum^{B-1}_{b=0} \frac{4}{\Tb} \Re\left[\sum_{k=\Ks}^{\Ke} w^{(b)}(f^{(b)}_k) \tilde{D}^{(b)\ast}_k \tilde{h}(f^{(b)}_k) \right]. \label{eq:d_h}
\end{aligned}
\end{equation}
$\tilde{D}^{(b)}_k$ can be computed from \eqref{eq:Dbm} and \eqref{eq:tDbm}, and stored before the sampling.
The frequency interval of the $b$-th band is $1/T^{(b)}$, which is larger than the original frequency interval of $1/T$ for $b\geq1$.
It means \eqref{eq:d_h} requires fewer waveform evaluations for $b\geq1$.
The number of waveform evaluations is
\begin{equation}
\KMB = \sum_{b=0}^{B-1} \left(\Ke - \Ks + 1\right),
\end{equation}
and its cost is reduced by a factor of $\Korig / \KMB$.
Since the computation of \eqref{eq:d_h} does not require the upsampling of $\tilde{h}(f^{(b)}_k)$, the computation of $(\bm{d}, \bm{h})$ is sped up by the same factor.

\subsection{How to compute $(h, h)$} \label{sec:h_h}

Next, we introduce approximate methods to compute $(\bm{h}, \bm{h})$.
It can be rewritten as follows thanks to \eqref{eq:unity},
\begin{equation}
(\bm{h}, \bm{h}) = \sum^{B-1}_{b=0} \frac{4}{T} \sum_{k=1}^{\floor{(\Nb-1)/2}} w^{(b)}(f_k) \frac{\left| \tilde{h}(f_k) \right|^2}{S_k}. \label{eq:h_h_original}
\end{equation}
As shown in the following, it can be approximately computed with waveforms at $f^{(b)}_k~(b=0, 1, \dots, B-1; k=\Ks,\Ks+1,\dots,\Ke)$, which have been computed for $(\bm{d},\bm{h})$.

\begin{figure}[t]
        \begin{center}
                \includegraphics[width = \columnwidth]{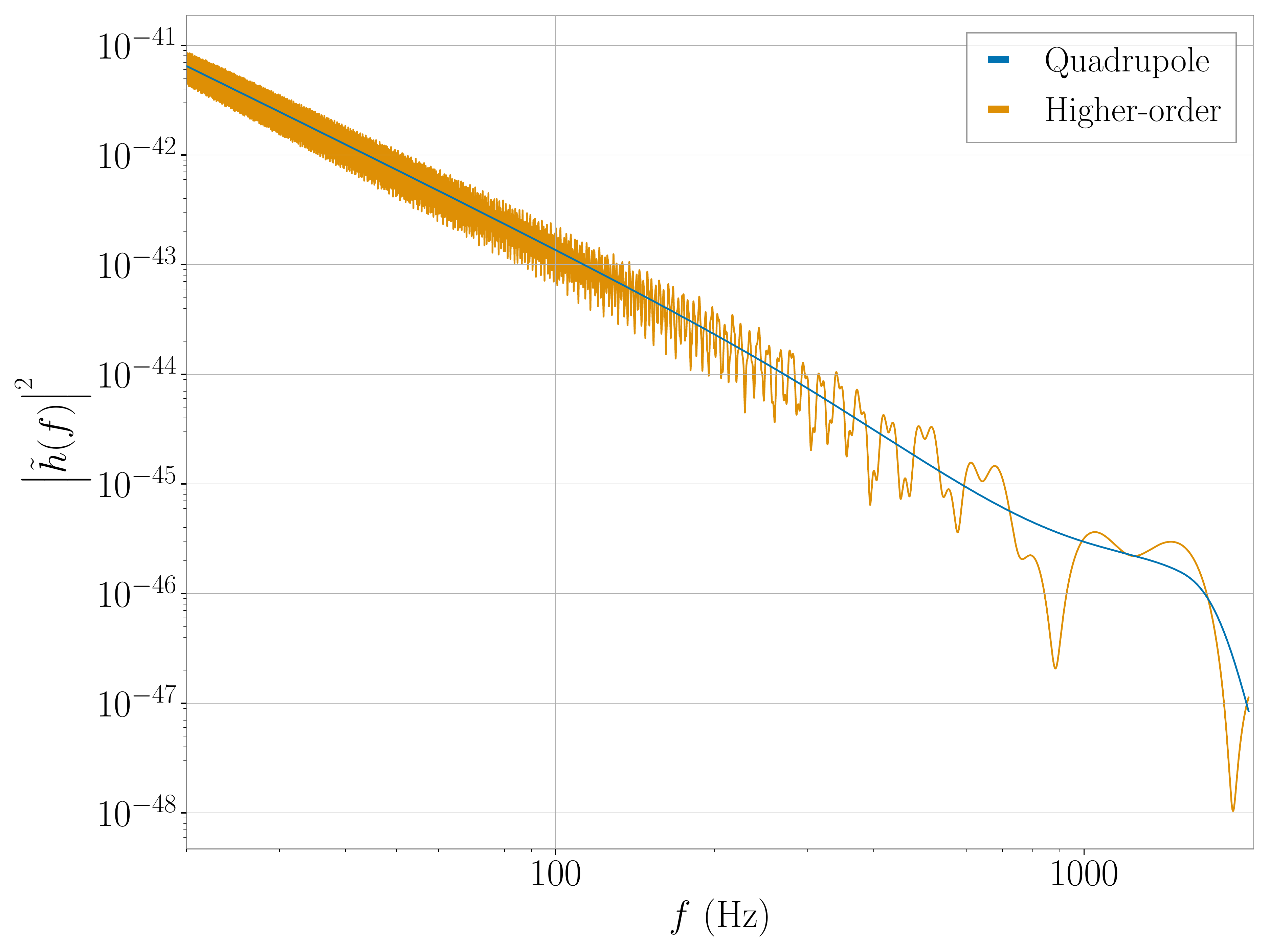}
                \caption{$\left| \tilde{h}(f) \right|^2$ for waveform models containing only dominant quadrupole moments (blue) and containing higher-order moments (orange).
                They are for non-spinning $7\Msun\text{--}1.4\Msun$ binary, whose inclination angle between the line of sight and the orbital angular momentum is $\pi/2$.
                The waveform models are IMRPhenomD \cite{Khan:2015jqa} for the quadrupole case and IMRPhenomHM \cite{London:2017bcn} for the higher-order case. }
                \label{fig:hsquare}
        \end{center}
\end{figure}

The waveform of CBC signal can be expressed as the linear combination of the $-2$ spin-weighted spherical harmonics ${}_{-2} Y_{lm}(\theta, \phi)$ \cite{Thorne:1980ru}.
The dominant moments are the quadrupole moments $(l, m)=(2, \pm2)$, and moments with $|m|\geq3$ are referred to as higher-order moments.
Figure \ref{fig:hsquare} shows $\left| \tilde{h}(f) \right|^2$ for waveform models containing only dominant quadrupole moments and containing higher-order moments.
In the former case, $|\tilde{h}(f)|^2$ is a smooth function as the phase is canceled out.
In the latter case, the cross terms between different moments give rise to an oscillatory behavior.
To compute $(\bm{h}, \bm{h})$ efficiently and accurately in each case, we propose two methods: {\it Linear interpolation} and {\it IFFT-FFT}.
The former method is more efficient but may not be accurate for waveform models containing higher-order moments.
The latter method is accurate for such waveform models but more costly.

\subsubsection{Linear interpolation}

In this method, $\left| \tilde{h}(f) \right|^2$ is approximated by the linear interpolation of $|\tilde{h}(f^{(b)}_k)|^2$,
\begin{equation}
\begin{aligned}
\left| \tilde{h}(f) \right|^2 &\simeq T^{(b)} \left(f^{(b)}_{k+1} - f\right) \left| \tilde{h}(f^{(b)}_k) \right|^2  \\
&~~~~~+ T^{(b)} \left(f - f^{(b)}_k\right) \left| \tilde{h}(f^{(b)}_{k+1}) \right|^2,
\end{aligned}
\end{equation}
for $f^{(b)}_k \leq f < f^{(b)}_{k+1}$ when the inner product in the $b$-th band is computed.
The approximate forms in $f^{(b)}_{\Ks} \leq f < f^{(b)}_{\Ks+1}$ and $f^{(b)}_{\Ke - 1} \leq f < f^{(b)}_{\Ke}$ are extrapolated to $f < f^{(b)}_{\Ks}$ and $f \geq f^{(b)}_{\Ke}$ respectively.

Substituting the linear interpolation into \eqref{eq:h_h_original}, we obtain
\begin{equation}
(\bm{h}, \bm{h}) = \sum^{B-1}_{b=0} \sum_{k=\Ks}^{\Ke} c^{(b)}_k \left| \tilde{h}(f^{(b)}_k) \right|^2. \label{eq:linear_interpolation}
\end{equation}
The coefficients are given by
\begin{equation}
\begin{aligned}
c^{(b)}_k &= \frac{4 T^{(b)}}{T} \sum_{\bar{f}^{(b)}_{k} \leq f_l < \bar{f}^{(b)}_{k+1}} \left(f^{(b)}_{k+1} - f_l\right) \frac{w^{(b)}(f_l)}{S(f_l)} \\
&~~ + \frac{4 T^{(b)}}{T} \sum_{\bar{f}^{(b)}_{k-1} \leq f_l < \bar{f}^{(b)}_{k}} \left(f_{l} - f^{(b)}_{k-1}\right) \frac{w^{(b)}(f_l)}{S(f_l)},
\end{aligned}
\end{equation}
where
\begin{equation}
\begin{aligned}
&\bar{f}^{(b)}_{\Ks - 1} = \bar{f}^{(b)}_{\Ks} = 0,~~~\bar{f}^{(b)}_{\Ke} = \bar{f}^{(b)}_{\Ke+1} = \infty,\\
&\bar{f}^{(b)}_{k} = f^{(b)}_k.~(k=\Ks+1,\Ks+2,\dots,\Ke-1)
\end{aligned}
\end{equation}
The computation of \eqref{eq:linear_interpolation} only requires $\mathcal{O}(\KMB)$ floating-point operations, and its cost is negligible compared to that of the waveform evaluations. 
The interpolation may not be accurate if the waveform model takes into account higher-order moments due to the oscillatory behavior of $\left| \tilde{h}(f) \right|^2$.

\subsubsection{IFFT-FFT}

The approximation used in this method is similar to that used for computing $(\bm{d}, \bm{h})$.
First, the inner product can be transformed into a sum over times as follows,
\begin{equation}
\begin{aligned}
&\frac{4}{T} \sum_{k=1}^{\floor{(\Nb-1)/2}} \wb(f_k) \frac{\left| \tilde{h}(f_k) \right|^2}{S_k} \\
&= 2 \dtb \sum^{\Nb - 1}_{m=0} I^{(b)}_m H^{(b)}_m, \label{eq:h_h_td}
\end{aligned}
\end{equation}
where
\begin{align}
&I^{(b)}_m \equiv \frac{2}{T} \Re \left[\sum_{k=1}^{\floor{(\Nb-1)/2}} \frac{1}{S_k} \e^{2 \pi \iu k m / \Nb} \right], \label{eq:I_def} \\
&H^{(b)}_m \equiv \nonumber \\
&~~\frac{2}{T} \Re \left[\sum_{k=1}^{\floor{(\Nb-1)/2}}  \wb(f_k) \left| \tilde{h}(f_k) \right|^2 \e^{2 \pi \iu k m / \Nb} \right].
\end{align}
$H^{(b)}_m$ is the convolution of the windowed waveform,
\begin{equation}
H^{(b)}_m = \dt^{(b)} \sum_{m'=0}^{N^{(b)}-1} \hat{h}^{(b)}_{\mod(m+m',\,\Nb)} \hat{h}^{(b)}_{m'},
\end{equation}
where $\mod(a,\,b)$ is the remainder of $a$ by $b$, and
\begin{equation}
\begin{aligned}
&\hat{h}^{(b)}_m \equiv \frac{2}{T} \times \\
&~~\Re \left[\sum_{k=1}^{\floor{(\Nb-1)/2}} \sqrt{\wb(f_k)} \tilde{h}(f_k) \e^{2 \pi \iu k m / \Nb} \right]. 
\end{aligned}
\end{equation}

The same argument in \secref{sec:d_h} leads to,
\begin{equation}
\hat{h}^{(b)}_m \simeq 0.~(m=0,1,\dots,\Nb - \Mb -1)
\end{equation}
If $2 \Mb - 1 < \Nb$, it means $H^{(b)}_m$ is vanishing for $\Mb \leq m \leq \Nb - \Mb$.
Thus, we can make the following approximation,
\begin{equation}
\sum^{\Nb - 1}_{m=0} I^{(b)}_m H^{(b)}_m \simeq \sum^{\hat{N}^{(b)} - 1}_{m=0} I^{(b)}_{\mathrm{c}, m} H^{(b)}_{\mathrm{c}, m}, \label{eq:h_h_approx1}
\end{equation}
where $\hat{N}^{(b)} \equiv \min\left[2 M^{(b)}, N^{(b)}\right]$, and $I^{(b)}_{\mathrm{c}, m}$ and $H^{(b)}_{\mathrm{c}, m}$ are the cropped sequences with the sizes of $\hat{N}^{(b)}$,
\begin{align}
&I^{(b)}_{\mathrm{c}, m} \equiv \begin{cases}
I^{(b)}_m, & (m \leq \floor{\hat{N}^{(b)} / 2}) \\
I^{(b)}_{m+\Nb-\hat{N}^{(b)}}, & (m \geq \floor{\hat{N}^{(b)} / 2}+1)
\end{cases} \label{eq:It_def} \\
&H^{(b)}_{\mathrm{c}, m} \equiv \begin{cases}
H^{(b)}_m, & (m \leq \floor{\hat{N}^{(b)} / 2}) \\
H^{(b)}_{m+\Nb-\hat{N}^{(b)}}. & (m \geq \floor{\hat{N}^{(b)} / 2}+1)
\end{cases} 
\end{align}
$H^{(b)}_{\mathrm{c}, m}$ can be expressed as the convolution of the cropped waveform,
\begin{equation}
H^{(b)}_{\mathrm{c}, m} = \dtb \sum_{m'=0}^{\hat{N}^{(b)}-1} \hat{h}^{(b)}_{\mathrm{c}, \mod(m+m',\,\hat{N}^{(b)})} \hat{h}^{(b)}_{\mathrm{c}, m'},
\end{equation}
where
\begin{equation}
\hat{h}^{(b)}_{\mathrm{c}, m} \equiv \hat{h}^{(b)}_{m+\Nb-\hat{N}^{(b)}}.~~(m=0,1, \dots,\hat{N}^{(b)}-1)
\end{equation}
Using the properties of the Fourier transformation, we obtain
\begin{equation}
\begin{aligned}
&2 \dtb \sum^{\hat{N}^{(b)} - 1}_{m=0} I^{(b)}_{\mathrm{c}, m} H^{(b)}_{\mathrm{c}, m} \\
&= \frac{4}{\hat{T}^{(b)}} \sum^{\floor{(\hat{N}^{(b)}-1)/2}}_{k=1} \tilde{I}^{(b)}_{\mathrm{c}, k}  \left|\tilde{h}^{(b)}_{\mathrm{c}, k} \right|^2, \label{eq:h_h_approx2}
\end{aligned}
\end{equation}
where $\hat{T}^{(b)}\equiv \hat{N}^{(b)} \dt^{(b)}$ and
\begin{align}
&\tilde{I}^{(b)}_{\mathrm{c}, k} = \dtb \sum_{m=0}^{\hat{N}^{(b)}-1} I^{(b)}_{\mathrm{c}, m} \e^{-2 \pi \iu k m / \hat{N}^{(b)}}, \label{eq:tIt_def} \\
&\tilde{h}^{(b)}_{\mathrm{c}, k} = \dtb \sum_{m=0}^{\hat{N}^{(b)}-1} h^{(b)}_{\mathrm{c}, m} \e^{-2 \pi \iu k m / \hat{N}^{(b)}}.
\end{align}

Substituting \eqref{eq:h_h_td}, \eqref{eq:h_h_approx1} and \eqref{eq:h_h_approx2} into \eqref{eq:h_h_original} leads to
\begin{equation}
(\bm{h}, \bm{h}) \simeq \sum^{B-1}_{b=0} \frac{4}{\hat{T}^{(b)}} \sum^{\floor{(\hat{N}^{(b)}-1)/2}}_{k=1} \tilde{I}^{(b)}_{\mathrm{c}, k} \left|\tilde{h}^{(b)}_{\mathrm{c}, k} \right|^2. \label{eq:h_h_HoM}
\end{equation}
$\tilde{I}^{(b)}_{\mathrm{c}, k}$ can be computed from \eqref{eq:I_def}, \eqref{eq:It_def} and \eqref{eq:tIt_def}, and stored before the sampling. 
$\tilde{h}^{(b)}_{\mathrm{c}, k}$ can be approximately computed as follows.
First, the last $\Mb$ components of $\hat{h}^{(b)}_{\mathrm{c},m}$ are computed as the inverse Fourier transform of the windowed waveform,
\begin{equation}
\begin{aligned}
&\hat{h}^{(b)}_{\mathrm{c},m} \simeq \frac{2}{\Tb} \times \\
&~~\Re\Bigg[\sum^{\floor{(\Mb - 1)/2}}_{k=1} \sqrt{\wb(f^{(b)}_k)} \tilde{h}(f^{(b)}_k) \e^{2 \pi \iu k m / \Mb} \Bigg], \\
&(m=\hat{N}^{(b)} - \Mb, \hat{N}^{(b)} - \Mb + 1, \dots, \hat{N}^{(b)}-1)
\end{aligned}
\end{equation}
and the first $\hat{N}^{(b)} - \Mb$ components of $\hat{h}^{(b)}_{\mathrm{c},m}$ are set to be zeros.
Then, $\tilde{h}^{(b)}_{\mathrm{c}, k}$ can be computed as the Fourier transform of $\hat{h}^{(b)}_{\mathrm{c}, m}$.
Thus, the computation of $\tilde{h}^{(b)}_{\mathrm{c}, k}$ requires an inverse fast Fourier transform (IFFT) and a fast Fourier transform (FFT), and this method is more costly than the linear interpolation method.
This IFFT-FFT operation requires $\mathcal{O}(\hat{N}^{(b)} \log_2 \hat{N}^{(b)})$ floating-point operations for each band.
Since $\hat{N}^{(b)} \leq 2M^{(b)} \sim 4 f^{(b+1)} T^{(b)}$, $2 f^{(b+1)} \ll 1 / \dt$ for small $b$, and $T^{(b)} \ll T$ for large $b$, we have $\hat{N}^{(b)} \ll N$.
Thus, unless we have a lot of redundant bands with similar values of $T^{(b)}$, the IFFT-FFT operations do not cause a fixed cost of $\mathcal{O}(N)$.

\subsection{How to determine the frequency bands $\{\fb\}_{b=0}^{B}$}

We assume $\{\Tb\}_{b=0}^{B-1}$ are specified by the user.
Then, the frequency bands $\{\fb\}_{b=0}^{B}$ should be determined so that the windowed waveform in the $b$-th band is vanishing at $t<T - \Tb$, where the start time of data is $t=0$.

The asymptotic behavior of the windowed waveform is studied in \appref{sec:choice_of_window}.
If we choose the following value as $\dfb$,
\begin{equation}
\dfb = \frac{1}{\sqrt{- \tau'(\fb)}}, \label{eq:dfb}
\end{equation}
the windowed waveforms behave as follows,
\begin{align}
&h^{(b)}_m \propto \left(\frac{\sqrt{-\tau'(\fb)}}{\tc - \tau(\fb) - m \dtb} \right)^3, \\
&\hat{h}^{(b)}_m \propto \left(\frac{\sqrt{-\tau'(\fb)}}{\tc - \tau(\fb) - m \dtb} \right)^2,
\end{align}
where $\tc$ is the time at which the coalescence part of signal arrives at the detector, according to \eqref{eq:smooth_window_behavior} and \eqref{eq:square_window}.
Thus, $\fb$ should satisfy
\begin{equation}
\frac{\sqrt{-\tau'(\fb)}}{\tc - \tau(\fb) - T + \Tb} \ll 1.
\end{equation}
For this condition to be satisfied, $\fb$ is determined by the following equation,
\begin{equation}
\tau(\fb) + L \sqrt{-\tau'(\fb)} = \Tb + \tcmin - T, \label{eq:band_equation}
\end{equation}
where $L$ is a user-specified constant satisfying $L \gg 1$, and $\tcmin$ is the minimum of $\tc$.
$L$ refers to the duration of the tail part of the windowed waveform taken in the segment $\Tb$ per $\sqrt{-\tau'(\fb)}$.
Larger $L$ increases the accuracy of the approximation, as the power of neglected part of waveform becomes less.
As shown in \secref{sec:validation}, $L=5$ is large enough for signals with signal-to-noise ratios of $\sim 25$.
Given $\fb$ satisfying \eqref{eq:band_equation}, $\dfb$ is determined by \eqref{eq:dfb}.

$\tcmin$ is determined by the prior range of $t_{\bigoplus}$, which is the time at which the coalescence part of signal arrives at the geocenter, and the light-traveling time from the geocenter to the detector.
We use the following conservative estimate,
\begin{equation}
\tcmin = t_{\bigoplus, \mathrm{min}} - \frac{R_{\bigoplus}}{c}, \label{eq:tcmin}
\end{equation}
where $t_{\bigoplus, \mathrm{min}}$ is the minimum of $t_{\bigoplus}$ in the prior range, $R_{\bigoplus}$ is the radius of the Earth, and $c$ is the light speed.
In the standard parameter estimation performed by the LIGO-Virgo collaboration, $t_{\bigoplus, \mathrm{min}}=T-2.1\,\si{\second}$ \cite{Veitch:2014wba}, and hence $\tcmin-T=-2.12\,\si{\second}$.
We use that standard prior range of $t_{\bigoplus, \mathrm{min}}$ and that value of $\tcmin$ throughout this paper unless specified otherwise.

For $\tau(f)$, we use the following leading-order expression in the Post-Newtonian (PN) expansion \cite{Blanchet:2006zz},
\begin{equation}
\tau_{0\mathrm{PN}}(f) = \frac{5}{256} \frac{G \Mc}{c^3} \left(\frac{\pi G \Mc f}{c^3}\right)^{-\frac{8}{3}},
\end{equation}
if the waveform contains only dominant quadrupole moments.
$G$ is the gravitational constant, and $\Mc$ is so-called chirp mass defined by
\begin{equation}
\Mc = \frac{(m_1 m_2)^{\frac{3}{5}}}{(m_1 + m_2)^{\frac{1}{5}}},
\end{equation}
where $m_1$ and $m_2$ are the masses of colliding objects.
We compute $\tau(f)$ with the minimum of chirp mass in the prior range, which gives the most conservative estimates.
If the waveform model takes into account higher-order moments, we use
\begin{equation}
\tau(f) = \tau_{0\mathrm{PN}}\left(\frac{2}{m_{\mathrm{max}}} f\right), \label{eq:tau_HoM}
\end{equation}
where $\mmax$ is the maximum of the magnetic numbers of moments for conservative estimates.

\subsection{Choice of $\df^{(B)}$} \label{sec:dfB}

\begin{figure}[t]
        \begin{center}
                \includegraphics[width = \columnwidth]{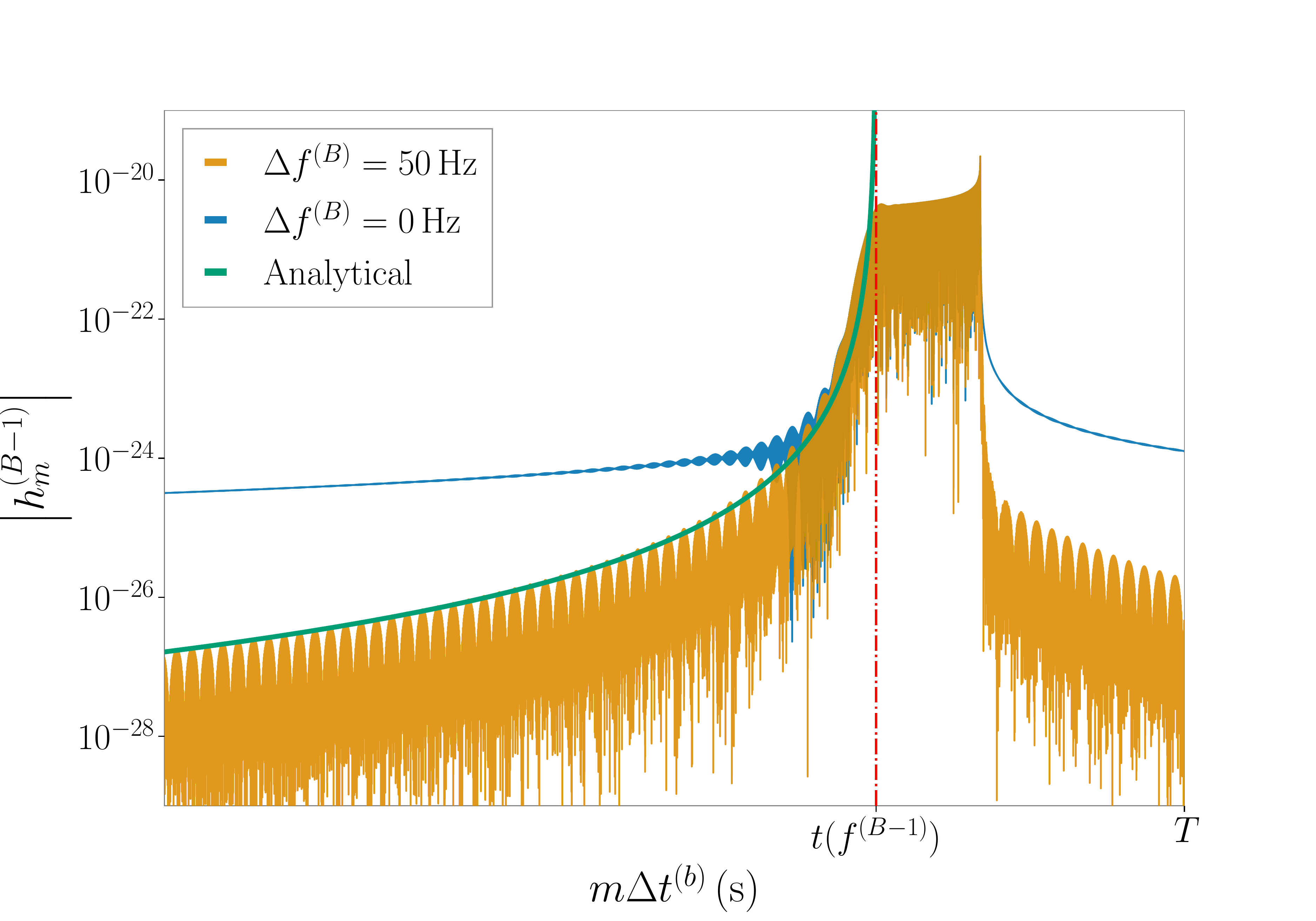}
                \caption{The inverse Fourier transforms of the waveforms in the $(B-1)$-th frequency band for $\df^{(B)}=0\,\si{\hertz}$ and $\df^{(B)}=50\,\si{\hertz}$.
                The waveforms are for non-spinning $1.4\Msun\text{--}1.4\Msun$ BNS signal, whose coalescence time is at $T-2\,\si{\second}$.
                The high-frequency cutoff is $f^{(B)} - \df^{(B)}=2048\,\si{\hertz}$.
                The red dashed-dotted line represents the time at which the frequency of the waveform is $f^{(B-1)}$, where $f^{(B-1)}=125.7\,\si{\hertz}$ for this plot.
                The green line represents the analytical prediction given by the right-hand side of \eqref{eq:smooth_window_behavior}.}
                \label{fig:highend}
        \end{center}
\end{figure}

$\df^{(B)}$ needs to be positive to smooth the high-frequency end of waveform.
Figure \ref{fig:highend} shows the absolute value of $h^{(B-1)}_m$ for $\df^{(B)}=0\,\si{\hertz}$ and $\df^{(B)}=50\,\si{\hertz}$, in comparison with the analytical prediction given by the right-hand side of \eqref{eq:smooth_window_behavior}.
For $\df^{(B)}=0\,\si{\hertz}$, $h^{(B-1)}_m$ has a long tail, which does not decay following the analytical prediction, due to the abrupt cutoff at the high-frequency end.
It significantly degrades the accuracy of our approximation.
For $\df^{(B)}=50\,\si{\hertz}$, $h^{(B-1)}_m$ quickly decays following the analytical prediction.

Assuming that the smoothed waveform decays with the timescale of $1/\df^{(B)}$, we use the following value,
\begin{equation}
\df^{(B)} = \frac{100}{T - \tcmax},
\end{equation}
where $\tcmax$ is the maximum of $\tc$, so that $h^{(b)}_m$ rapidly decays in the last $T - \tcmax$ of data.
Following the same argument for deriving \eqref{eq:tcmin}, we use the following conservative value of $\tcmax$,
\begin{equation}
\tcmax = t_{\bigoplus, \mathrm{max}} + \frac{R_{\bigoplus}}{c},
\end{equation}
where $t_{\bigoplus, \mathrm{max}}$ is the maximum of $t_{\bigoplus}$ in the prior range.
In the standard parameter estimation performed by the LIGO-Virgo collaboration, $t_{\bigoplus, \mathrm{max}}=T-1.9\,\si{\second}$ \cite{Veitch:2014wba}, which leads to $\df^{(B)}\simeq 53\,\si{\hertz}$.
We use that standard value throughout this paper unless specified otherwise.

\subsection{Speed-up gains} \label{sec:speedup}

\begin{table*}[t]
\setlength{\tabcolsep}{9pt}
\centering
\begin{tabular}{c | c | c | c c | c c}

\multirow{2}{*}{$\flow~(\si{\hertz})$} &
\multirow{2}{*}{$T~(\si{\second})$} &
\multirow{2}{*}{$\Korig$} &
\multicolumn{2}{c |}{IMRPhenomD} &
\multicolumn{2}{c}{IMRPhenomHM}  \\ 
  
  & & &
$\Korig/\KMB$ \!& Speed up \!&
$\Korig/\KMB$ \!& Speed up \\ \hline

$20$ & 256 & $5.2 \times 10^5$ &
$4.5\times10$ & $5.1\times10$ &
$2.7\times10$ & $2.1\times10$ \\ \hline

$10$ & $1024$ & $2.1 \times 10^6$ &
$1.2\times10^2$ & $1.5\times10^2$ &
$5.7\times10$ & $4.6\times10$ \\ \hline

$5$ & $8192$ & $1.7 \times 10^7$ &
$4.4\times10^2$ & $4.9\times10^2$ &
$1.6\times10^2$ & $1.2\times10^2$ \\

\end{tabular}
\caption{$\Korig/\KMB$ and speed-up gains in evaluations of log-likelihood-ratio for non-spinning $1.4\Msun$--$1.4\Msun$ BNS.
The table lists their values for various values of low-frequency cutoffs $\flow$ and durations $T$, and the IMRPhenomD and IMRPhenomHM waveform models.
The high-frequency cutoff is $2048\,\si{\hertz}$.
The speed-up gains were measured on 8-core Intel Core i9 with the clock rate of $2.4\,\si{\giga \hertz}$.}
\label{tab:speed_up}
\end{table*}

Finally, we evaluate the speed-up gains of our technique.
We implemented our technique based on the likelihood class of BILBY \cite{Ashton:2018jfp, Romero-Shaw:2020owr}, and measured the speed-up gains in evaluations of log-likelihood-ratio.
Since the run time of parameter estimation is approximately the product of the evaluation time of log-likelihood-ratio and the number of their evaluations, it approximates the overall speed-up gain in parameter estimation.

Table \ref{tab:speed_up} shows $\Korig/\KMB$ and speed-up gains for non-spinning $1.4\Msun$--$1.4\Msun$ BNS.
The table lists their values for various values of $\flow$, and $T$ is chosen as the minimum power of $2$ larger than the time-to-merger of dominant quadrupole moments from $\flow$.
It effectively sets higher low-frequency cutoffs on higher-order moments.
The high-frequency cutoff is $2048\,\si{\hertz}$.
The total frequency range is divided into frequency bands determined by \eqref{eq:band_equation} with $\{\Tb\}_{b=0}^{B-1}=\{T,~T/2,~T/4,~\cdots,~4\,\si{\second}\}$ and $L=5$.
IMRPhenomD \cite{Khan:2015jqa} and IMRPhenomHM \cite{London:2017bcn} are chosen as representative waveform models.
IMRPhenomHM includes the effects of higher-order multiple moments, and $\mmax=4$ is used to compute $\tau(f)$ when the frequency bands are calculated.
For the computation of $(\bm{h}, \bm{h})$, the linear-interpolation method and the IFFT-FFT method were used for IMRPhenomD and IMRPhenomHM respectively.

The speed-up gains are roughly equal to $\Korig/\KMB$, which means our method does not suffer from the fixed cost present in MB-Interpolation \cite{Vinciguerra:2017ngf}.
The speed-up gains are smaller for IMRPhenomHM, as higher-order multiple moments have longer time-to-merger from a given frequency, and \eqref{eq:band_equation} gives more severe constraints.
For $\flow=20\,\si{\hertz}$, which is used in most of the analyses by the LIGO-Virgo collaboration \cite{LIGOScientific:2018mvr, Abbott:2020niy}, the speed-up gains are $\mathcal{O}(10)$ in both cases.
For $\flow=5\,\si{\hertz}$, which can be used for the third-generation detectors with improved sensitivities at low frequencies \cite{Hild:2010id, Evans:2016mbw}, the speed-up gains are $\mathcal{O}(10^2)$.

\section{Validation} \label{sec:validation}

In the previous section, we have formulated our technique, and shown that it significantly speeds up the parameter estimation.
In this section, we investigate the accuracy of our technique.

\subsection{Likelihood errors for GW190814}

First, we investigate the errors of log-likelihood-ratio $\ln \Lambda$ from our approximation for GW190814 \cite{Abbott:2020khf}, gravitational-wave signal detected by the LIGO-Virgo collaboration.
We computed $\ln \Lambda$ with and without our approximation on the posterior samples from the parameter estimation of this signal, and took their differences $\Delta \ln \Lambda$ as the errors of our approximation.
This signal has a relatively large signal-to-noise ratio (SNR) of $\sim 25$, and it is appropriate for our study as systematic errors become prominent for a large SNR.
This signal contains higher-order multipole moments at high confidence, which enables us to study the accuracy of our technique in their presence.
The data, PSD, and posterior samples were obtained from the Gravitational Wave Open Science Center \cite{gwosc}.

\begin{figure*}[t]
        \begin{center}
                \includegraphics[width = 0.49\linewidth]{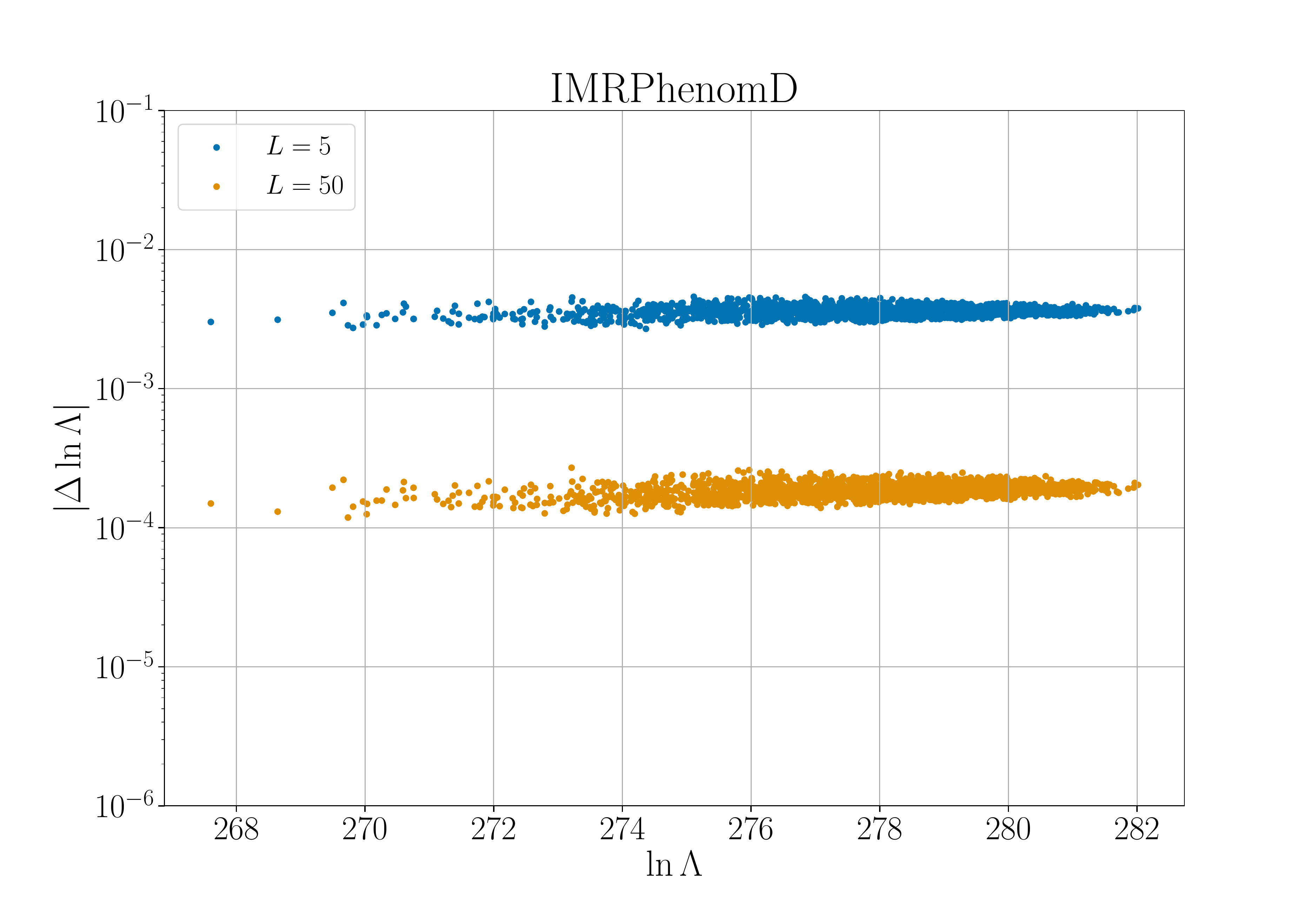}
                \includegraphics[width = 0.49\linewidth]{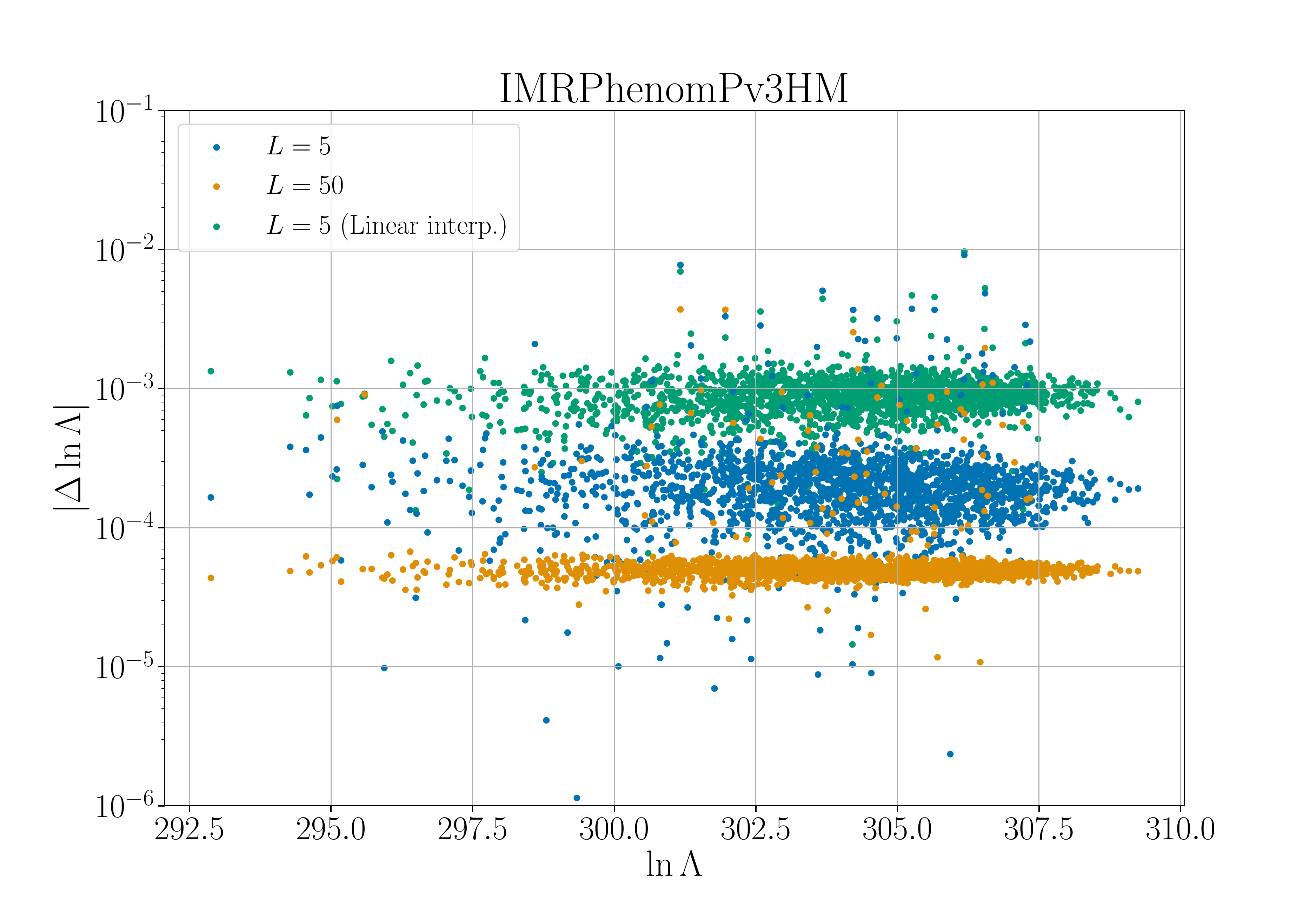}
                \caption{The errors of log-likelihood-ratio, $\ln \Lambda$, from our approximation for GW190814.
                The left figure shows the errors for the IMRPhenomD waveform model, and the right figure shows those for IMRPhenomHM.
                In each figure, the errors for $L=5$ and $L=50$ are shown in blue and orange respectively.
                For the computation of $(\bm{h}, \bm{h})$, the linear-interpolation method and the IFFT-FFT method are used for IMRPhenomD and IMRPhenomHM respectively by default.
                For IMRPhenomHM, the errors from the linear-interpolation method and $L=5$ are also shown in green.
                For visibility, the data points are downsampled to 2000.}
                \label{fig:likelihood_errors}
        \end{center}
\end{figure*}

We computed the errors for two different waveform models, IMRPhenomD and IMRPhenomPv3HM \cite{Khan:2018fmp, Khan:2019kot}.
For each waveform model, we used the posterior samples from the analysis using the same waveform model.
The results are shown in \figref{fig:likelihood_errors}.
We used $16\,\si{\second}$ of data around the time of detection.
Following \cite{Abbott:2020khf}, we analyzed the frequency range of $20$--$1024\,\si{\hertz}$ for LIGO-Hanford and Virgo, and $30$--$1024\,\si{\hertz}$ for LIGO-Livingston.
The total frequency range is divided into $3$ bands determined by \eqref{eq:band_equation} with $\{\Tb\}_{b=0}^{2}=\{16\,\si{\second},~8\,\si{\second},~4\,\si{\second}\}$, and $L=5$ or $L=50$.
For computing $\tau(f)$, we used the reference chirp mass of $6.4\Msun$, which is the median of the inferred detector-frame chirp mass, and $\mmax=4$ for IMRPhenomPv3HM.
We ignored the calibration errors of detectors as their effects are expected to be negligible for the SNR of this signal \cite{Payne:2020myg}.

For IMRPhenomD, we used the linear-interpolation method to compute $(\bm{h}, \bm{h})$.
The median error is $4 \times 10^{-3}$ for $L=5$ and $2 \times 10^{-4}$ for $L=50$, which shows increasing $L$ improves the accuracy.
The number of waveform evaluations is reduced by a factor of $3.6$ for $L=5$ and $3.2$ for $L=50$.
For IMRPhenomPv3HM, we primarily used the IFFT-FFT method to compute $(\bm{h}, \bm{h})$.
The median error is $2 \times 10^{-4}$ for $L=5$ and $5 \times 10^{-5}$ for $L=50$, which again shows increasing $L$ improves the accuracy.
The number of waveform evaluations is reduced by a factor of $3.3$ for $L=5$ and $2.8$ for $L=50$.
For IMRPhenomPv3HM, the figure also shows the errors from the linear-interpolation method and $L=5$.
These errors are larger than those with the IFFT-FFT method, but they are still well below unity.
In any case, the systematic errors due to our approximation are well below unity and the statistical errors.

\subsection{Consistency of parameter estimation}

To investigate the consistency of our technique, we performed parameter estimation of hundreds of simulated CBC signals using our technique.
For each signal, we constructed the credible interval of each parameter centered on its median, and computed the credible level at which its true value is found.
For the inference to be consistent, the credible levels should be uniformly distributed from 0 to 1 \cite{doi:10.1198/106186006X136976, 2018arXiv180406788T}.

We considered the network of the two LIGO detectors and the Virgo detector, and signals were injected into Gaussian noise colored by their design sensitivities.
We simulated 256 non-spinning BNS signals, whose chirp masses and mass ratios $q \equiv m_2 / m_1$ are distributed uniformly within
\begin{equation}
1.15\Msun \leq \Mc \leq 1.25\Msun,~~~~~~0.2 \leq q \leq 1.
\end{equation}
The luminosity distance $\DL$ ranges from $10\,\si{\mega \parsec}$ to $100\,\si{\mega \parsec}$, and its distribution is proportional to $\DL^2$.
The locations of the sources and the directions of the orbital angular momenta are isotropically distributed.
The waveform model of simulated signals is IMRPhenomD, and the same waveform model was used for parameter estimation.
The median network SNR of simulated signals is $24.3$.

For parameter estimation, we used BILBY as an interface between likelihood and sampler, and DYNESTY \cite{Speagle:2019ivv} as sampler.
The prior is the same as the distribution of simulated signals.
The coalescence phase was analytically marginalized over and the luminosity distance was marginalized over with the look-up table method \cite{Singer:2015ema, Thrane:2018qnx}.
The total frequency range is $20$--$2048\,\si{\hertz}$, and it is divided into $7$ bands determined by \eqref{eq:band_equation} with $\{\Tb\}_{b=0}^{6}=\{256\,\si{\second},~128\,\si{\second},~\cdots,~4\,\si{\second}\}$, $L=5$ and the reference chirp mass of $1.15\Msun$.
The number of waveform evaluations is reduced by a factor of $44$.

\begin{figure}[t]
        \begin{center}
                \includegraphics[width = 0.85\columnwidth]{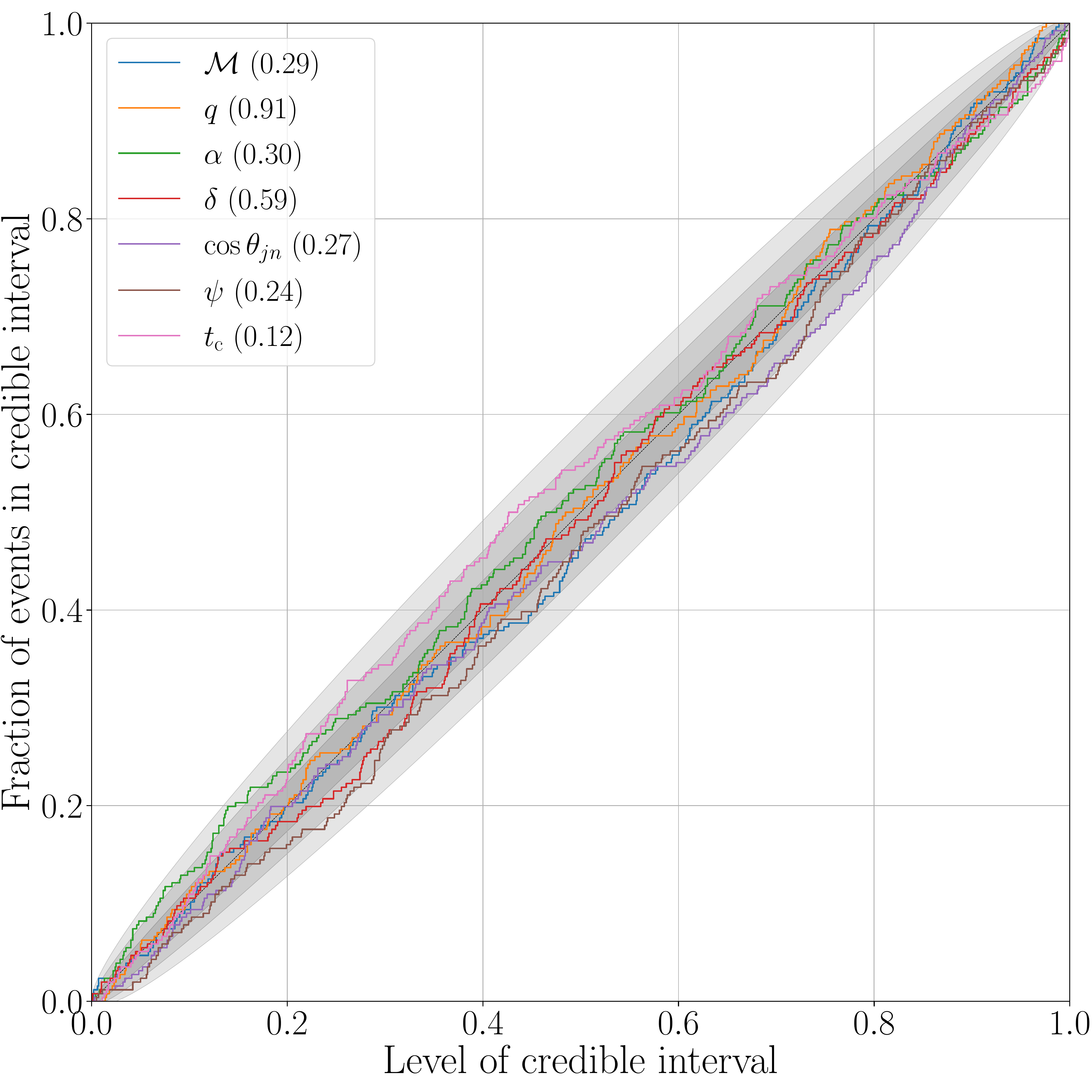}
                \caption{The cumulative distribution of credible levels for each source parameter, obtained from $256$ simulated CBC signals. 
                The gray regions represent the $1\text{--}\sigma$, $2\text{--}\sigma$ and $3\text{--}\sigma$ confidence intervals of statistical errors due to a finite number of samples.
                Each label shows the $p$-value of the Kolmogorov-Smirnov test between the credible levels and a uniform distribution.}
                \label{fig:p_p_plot}
        \end{center}
\end{figure}

Figure \ref{fig:p_p_plot} shows the cumulative distribution of credible levels for each parameter.
If credible levels are uniformly distributed, they should be diagonal lines for an infinite number of samples.
The gray regions represent the $1\text{--}\sigma$, $2\text{--}\sigma$ and $3\text{--}\sigma$ confidence intervals of statistical errors due to a finite number of samples, and the distributions are inside the $3\text{--}\sigma$ interval for most of the range.
The figure also presents the $p$-values of Kolmogorov-Smirnov tests between the credible levels and a uniform distribution in the legend.
The moderate $p$-values indicate that the credible levels are consistent with uniformly distributed random numbers.

\section{Conclusion} \label{sec:conclusion}

In this paper, we have presented a technique to speed up the parameter estimation of gravitational waves from compact binary coalescence (CBC), which exploits the chirping behavior of CBC signal.
It does not require the upsampling of waveforms, which is required by the MB-Interpolation technique, another implementation of this idea proposed by \cite{Vinciguerra:2017ngf}.
Thus, our technique does not suffer from the fixed cost due to it, and the speed-up gains are larger.
In \secref{sec:speedup}, we have found that our technique speeds up the parameter estimation of $1.4\Msun$--$1.4\Msun$ binary neutron star (BNS) signal by a factor of $\mathcal{O}(10)$ for the low-frequency cutoff of $20\,\si{\hertz}$, which is the standard low-frequency cutoff used by the LIGO-Virgo analyses.
The speed-up gain is increased to $\mathcal{O}(10^2)$ for the low-frequency cutoff of $5\,\si{\hertz}$, which can be used for the third-generation detectors.
We have investigated the errors of log-likelihood-ratio from our approximation and the consistency of the inference using our technique in \secref{sec:validation}.
The results indicate our technique is accurate enough to be used for signals, which have relatively large signal-to-noise ratios (SNRs) of $\sim25$.
The errors of log-likelihood-ratio imply our technique is applicable to signals with higher SNRs.
Investigating the limitation on the accuracy of our technique is the future work.

We note that there are various other techniques proposed to reduce the cost of waveform evaluations in parameter estimation.
The reduced order quadrature (ROQ) technique \cite{Canizares:2014fya, Smith:2016qas, Morisaki:2020oqk} approximates waveforms by the linear combinations of basis vectors, and significantly reduces the number of frequency samples where waveforms are evaluated.
The speed-up gain for BNS signal with the low-frequency cutoff of $20\,\si{\hertz}$ is $\mathcal{O}(10^2)$ \cite{Canizares:2014fya, Smith:2016qas}, and it is increased to $\mathcal{O}(10^4)$ if basis vectors are constructed in narrow parameter space \cite{Morisaki:2020oqk}.
The heterodyned likelihood \cite{Cornish:2010kf, Cornish:2021wxy} and relative binning \cite{Zackay:2018qdy} methods assume waveforms sampled over in parameter estimation are very similar to the template waveform triggering the detection.
The speed-up gain of relative binning is $\mathcal{O}(10^4)$ for GW170817 \cite{Zackay:2018qdy}, BNS signal detected by the LIGO-Virgo collaboration.
Compared to those speed-up gains, the speed-up gain of our technique is more modest.
On the other hand, ROQ requires offline basis construction, which needs to be done for each waveform model we are interested in.
The heterodyned likelihood and relative binning methods require a reference waveform, which is very similar to the true waveform.
Since our technique does not require any offline preparations or reference waveforms, it is more easy-to-use than the other techniques.
We also note that our technique can be used to reduce the file size of ROQ basis vectors and speed up the pre-computations of ROQ, which is explained in \appref{sec:application_to_roq}.

\begin{acknowledgments}

The author thanks Rory Smith and Daniel Wysocki for helpful comments to improve this paper.
The author is supported by NSF PHY-1912649.
The author is grateful for computational resources provided by the Leonard E Parker Center for Gravitation, Cosmology and Astrophysics at the University of Wisconsin-Milwaukee and supported by National Science Foundation Grants PHY-1626190 and PHY-1700765.

This research has made use of data, software and/or web tools obtained from the Gravitational Wave Open Science Center (https://www.gw-openscience.org/), a service of LIGO Laboratory, the LIGO Scientific Collaboration and the Virgo Collaboration. LIGO Laboratory and Advanced LIGO are funded by the United States National Science Foundation (NSF) as well as the Science and Technology Facilities Council (STFC) of the United Kingdom, the Max-Planck-Society (MPS), and the State of Niedersachsen/Germany for support of the construction of Advanced LIGO and construction and operation of the GEO600 detector. Additional support for Advanced LIGO was provided by the Australian Research Council. Virgo is funded, through the European Gravitational Observatory (EGO), by the French Centre National de Recherche Scientifique (CNRS), the Italian Istituto Nazionale di Fisica Nucleare (INFN) and the Dutch Nikhef, with contributions by institutions from Belgium, Germany, Greece, Hungary, Ireland, Japan, Monaco, Poland, Portugal, Spain.

\end{acknowledgments}

\appendix

\section{Choice of window function} \label{sec:choice_of_window}

Our technique is based on the approximation that the windowed waveform in the $b$-th band is vanishing at $t \lesssim T - \tau(\fb - \dfb)$.
We investigate the asymptotic behavior of the windowed waveform at $t < T - \tau(\fb - \dfb)$ for various window functions.

\subsection{Stationary phase approximation}

Each moment of a CBC waveform can be modeled as follows,
\begin{equation}
h(t) = A(t) \cos \Phi(t),
\end{equation}
where $\Phi(t)$ is defined so that $\Phi'>0,~\Phi''>0$.
In the inspiral regime, the amplitude changes more slowly than the phase,
\begin{equation}
\left| \frac{A'}{A} \right| \ll \Phi',~~~~~~\Phi'' \ll \left(\Phi'\right)^2.
\end{equation}
In this regime, we can apply the stationary phase approximation to calculate the Fourier transform of $h(t)$ \cite{Creighton:2011zz},
\begin{align}
\tilde{h}(f) &= \int^\infty_{-\infty} h(t) \e^{- 2 \pi \iu f t} dt \\
&\simeq \begin{cases}
B(f) \e^{-\iu \Psi(f)}, & (f>0) \\
B(f) \e^{\iu \Psi(f)}, & (f<0)
\end{cases}
\end{align}
where
\begin{align}
B(f) &= \frac{\sqrt{t'(f)}}{2} A(t(f)), \\
\Psi(f) &= - \Phi(t(f)) + 2 \pi f t(f) - \frac{\pi}{4},
\end{align}
and $t(f)$ is the time at which $\Phi'(t) = 2 \pi f$.

\subsection{Rectangular window}

First, we consider the waveform windowed by a rectangular window,
\begin{equation}
h^{(b)}(t) = 2 \Re\left[ \int^{f^{(b+1)}}_{f^{(b)}} df B(f) \e^{2 \pi \iu f t - \iu \Psi(f)} \right].
\end{equation}
For $t<t(f^{(b)})$, most of the contributions to the integral come from around $f=f^{(b)}$.
Thus, we expand the integrand around $f=f^{(b)}$,
\begin{align}
&B(f) \simeq B(f^{(b)}), \label{eq:B_exp} \\
&\Psi(f) \simeq \Psi(f^{(b)}) + 2 \pi t(f^{(b)}) (f - f^{(b)}) \nonumber \\
&~~~~~~~~~~~~~~~~~~~~+ \pi t'(f^{(b)}) (f - f^{(b)})^2, \label{eq:Psi_exp}
\end{align}
and approximately evaluate the integral as follows,
\begin{align}
&h^{(b)}(t) \simeq 2 \Re \Bigg[ B(f^{(b)}) \e^{2 \pi \iu f^{(b)} t - \iu \Psi(f^{(b)})}  \nonumber \\
&~~\times \int_{f^{(b)}}^\infty df \e^{-2 \pi \iu (f - f^{(b)}) (t(f^{(b)}) - t) - \iu \pi t'(f^{(b)}) (f - f^{(b)})^2} \Bigg] \nonumber \\
&\simeq A(t(f^{(b)})) \Re \Bigg[ \e^{2 \pi \iu f^{(b)} t - \iu \Psi(f^{(b)})}  \nonumber \\
&\times \left(- \frac{\iu}{2 \pi} \frac{\sqrt{t'(f^{(b)})}}{t(f^{(b)}) - t} + \frac{1}{4 \pi^2} \left( \frac{\sqrt{t'(f^{(b)})}}{t(f^{(b)}) - t} \right)^3 \right) \Bigg], \label{eq:rectangular}
\end{align} 
where we have used
\begin{equation}
\int^{\infty}_0 \e^{-\iu p x - \iu q x^2} dx = - \frac{\iu}{p} + \frac{2q}{p^3} + \mathcal{O}\left(\frac{1}{p^5}\right),
\end{equation}
for $p>0$ and $q>0$.
\eqref{eq:rectangular} means $h^{(b)}(t)$ has a long tail inversely proportional to $t(f^{(b)}) - t$, which degrades the accuracy of our technique.

\subsection{Smooth window}

Next, we consider the smooth window given by \eqref{eq:window_def},
\begin{equation}
\begin{aligned}
&h^{(b)}(t) = \\
&~~2 \Re \left[ \int^{f^{(b+1)}}_{f^{(b)} - \df^{(b)}} df w^{(b)}(f) B(f) \e^{2 \pi \iu f t - \iu \Psi(f)} \right].
\end{aligned}
\end{equation}
Using the approximations, \eqref{eq:B_exp} and \eqref{eq:Psi_exp}, we can approximately evaluate the integral from $\fb - \dfb$ to $\fb$,
\begin{align}
&\int^{f^{(b)}}_{f^{(b)} - \df^{(b)}} df w^{(b)}(f) B(f) \e^{2 \pi \iu f t - \iu \Psi(f)}  \nonumber \\
&\simeq A(t(f^{(b)})) \e^{2 \pi \iu f^{(b)} t - \iu \Psi(f^{(b)})} \times  \nonumber \\
&\Bigg[ \frac{\iu}{4 \pi} \frac{\sqrt{t'(f^{(b)})}}{t(f^{(b)}) - t} - \frac{1}{8 \pi^2} \left(\frac{\sqrt{t'(f^{(b)})}}{t(f^{(b)}) - t}\right)^3  \nonumber \\
&~~~ + \frac{\iu}{32 \pi}  \frac{\sqrt{t'(f^{(b)})}}{\left(\df^{(b)} \right)^2 \left(t(f^{(b)}) - t\right)^3} \times \nonumber \\
&~~~~~ \left(1 + \e^{2 \pi \iu \df^{(b)} (t(f^{(b)}) - t) - \pi \iu t'(f^{(b)}) \left(\df^{(b)}\right)^2}\right)  \Bigg],
\end{align}
where we have used
\begin{align}
&\int^0_{-1} \left(1 + \cos(\pi x) \right) \e^{-\iu p x - \iu q x^2} \nonumber \\
&= \frac{2 \iu}{p} - \frac{4 q - \iu \pi^2 \left(1 + \e^{\iu (p - q)}\right) }{p^3} + \mathcal{O}\left( \frac{1}{p^4} \right).
\end{align}
The integral from $f^{(b)}$ to $f^{(b+1)}$ is the same as \eqref{eq:rectangular}, and the windowed waveform is given by
\begin{align}
&h^{(b)}(t) \simeq A(t(f^{(b)})) \times \nonumber \\
&\Re\Bigg[ \frac{\iu}{16 \pi}  \frac{\sqrt{t'(f^{(b)})}}{\left(\df^{(b)} \right)^2 \left(t(f^{(b)}) - t\right)^3}  \e^{2 \pi \iu f^{(b)} t - \iu \Psi(f^{(b)})} \nonumber \\
&~~~\times \left(1 + \e^{2 \pi \iu \df^{(b)} (t(f^{(b)}) - t) - \pi \iu t'(f^{(b)}) \left(\df^{(b)}\right)^2}\right) \Bigg]. \label{eq:smooth_window}
\end{align}
Its amplitude is quickly attenuated in proportion to $\left(t(f^{(b)}) - t\right)^{-3}$,
\begin{equation}
\left|h^{(b)}(t)\right| \leq \frac{A(t(f^{(b)}))}{8 \pi} \frac{\sqrt{t'(f^{(b)})}}{\left(\df^{(b)} \right)^2 \left(t(f^{(b)}) - t\right)^3}, \label{eq:smooth_window_behavior}
\end{equation}
and this smooth window is more appropriate than the rectangular window to be used for our technique.

\subsection{Square-root of smooth window}

Finally, we consider the square-root of the smooth window,
\begin{align}
&h^{(b)}(t) = \nonumber \\
&~~2 \Re \left[ \int^{f^{(b+1)}}_{f^{(b)} - \df^{(b)}} df \sqrt{\wb(f)} B(f) \e^{2 \pi \iu f t - \iu \Psi(f)} \right],
\end{align}
which is used in the IFFT-FFT method for the computation of $(\bm{h}, \bm{h})$.
The integral from $\fb - \dfb$ to $\fb$ can be approximately evaluated as follows,
\begin{align}
&\int^{f^{(b)}}_{f^{(b)} - \df^{(b)}} df \sqrt{w^{(b)}(f)} B(f) \e^{2 \pi \iu f t - \iu \Psi(f)} \nonumber \\
&\simeq A(t(f^{(b)})) \e^{2 \pi \iu f^{(b)} t - \iu \Psi(f^{(b)})}  \nonumber \\
& \times \Bigg[ \frac{\iu}{4 \pi} \frac{\sqrt{t'(f^{(b)})}}{t(f^{(b)}) - t} - \frac{1}{16 \pi} \frac{\sqrt{t'(\fb)}}{\dfb (t(\fb) - t)^2} \nonumber \\
&~~~~~ \times \e^{2 \pi \iu \dfb (t(\fb) - t) - \pi \iu t'(\fb) \left(\dfb\right)^2} \Bigg],
\end{align}
where the following formula has been used,
\begin{equation}
\int^0_{-1} \cos\left(\frac{\pi}{2} x\right) \e^{-\iu p x - \iu q x^2} dx= \frac{\iu}{p} - \frac{\pi  \e^{\iu(p-q)}}{2 p^2} + \mathcal{O}\left(\frac{1}{p^3}\right).
\end{equation}
Thus, the windowed waveform is approximately given by
\begin{align}
&h^{(b)}(t) \simeq - \frac{A(t(\fb))}{8 \pi} \frac{\sqrt{t'(\fb)}}{\dfb (t(\fb) - t)^2} \nonumber \\
&\times \cos\Bigg(2 \pi \fb t - \Psi(\fb) + 2 \pi \dfb (t(\fb) - t) \nonumber \\
&~~~~~~~~~~ - \pi t'(\fb) \left(\dfb\right)^2 \Bigg), \label{eq:square_window}
\end{align}
and it is quickly attenuated in proportion to $\left(t(f^{(b)}) - t\right)^{-2}$.

\section{Application to ROQ} \label{sec:application_to_roq}

Our technique can be used to reduce the file size of ROQ basis vectors and speed up the pre-computations of ROQ.
ROQ approximates template waveforms by the linear combinations of reduced basis vectors $\{B_j\}_{j=1}^{\NL}$ \cite{Canizares:2014fya, Smith:2016qas, Morisaki:2020oqk},
\begin{equation}
\tilde{h}(f_k) \simeq \sum_{j=1}^{\NL} B_j(f_k) \tilde{h}(F_j; \tc=0) \e^{- 2 \pi \iu f_k \tc}, \label{eq:roq_approx}
\end{equation}
where $\{F_j\}_{j=1}^{\NL}$ is the subset of frequency samples determined by the empirical interpolation algorithm (See the algorithm 2 of \cite{Field:2013cfa}).
Substituting it into the original form of $(\bm{d}, \bm{h})$, \eqref{eq:original_d_h}, we obtain
\begin{align}
&(\bm{d}, \bm{h}) \simeq \Re \left[ \sum_{j=1}^{\NL} \omega_j (\tc) \tilde{h}(F_j; \tc=0) \right], \\
&\omega_j (\tc) \equiv \frac{4}{T} \sum_{k=1}^{\floor{(N-1)/2}} \frac{\tilde{d}^{\ast}_k B_j(f_k)}{S_k} \e^{- 2 \pi \iu f_k \tc}. \label{eq:original_roq}
\end{align}
$\{\omega_j (\tc)\}_{j=1}^{\NL}$ are referred to as ROQ weights, and need to be pre-computed before sampling.
On the other hand, substituting \eqref{eq:roq_approx} into our approximate form of $(\bm{d}, \bm{h})$, \eqref{eq:d_h}, we obtain
\begin{align}
&(\bm{d}, \bm{h}) \simeq \Re \left[ \sum_{j=1}^{\NL} \omega^{\mathrm{MB}}_j (\tc) \tilde{h}(F_j; \tc=0) \right], \\
&\omega^{\mathrm{MB}}_j (\tc) \equiv \sum^{B-1}_{b=0} \frac{4}{\Tb} \times \nonumber \\
&~~\Re\left[\sum_{k=\Ks}^{\Ke} w^{(b)}(f^{(b)}_k) \tilde{D}^{(b)\ast}_k B_j (f^{(b)}_k) \e^{- 2 \pi \iu f^{(b)}_k \tc} \right]. \label{eq:mb_roq}
\end{align}

The computation of \eqref{eq:mb_roq} requires basis vectors only at $\KMB$ frequency samples while \eqref{eq:original_roq} requires them at all the $\Korig$ frequency samples.
This means we do not need to store basis vectors at all the frequency samples under our approximation, and their file size can be reduced by a factor of $\Korig/\KMB$.
Since the size of basis vectors for BNS waveforms can be $\mathcal{O}(10)\,\mathrm{GB}$ or even larger, this is practically useful.
Comparing \eqref{eq:original_roq} and \eqref{eq:mb_roq}, we also find that our technique reduces the floating-point operations required to calculate ROQ weights by a factor of $\Korig/\KMB$.

\bibliographystyle{apsrev4-1}
\bibliography{multibandedpe}

\end{document}